\pdfoutput=1
\documentclass[superscriptaddress, 12pt, aps, prl, preprint,nofootinbib]{revtex4-1}
\usepackage{graphicx}
\usepackage{epstopdf}
\usepackage{color}
\usepackage{amsmath, amssymb}

\begin{document}

\title{Large presence of carbonic acid in CO$_2$-rich aqueous fluids under Earth's mantle conditions}
\author{Nore Stolte}
\affiliation{Department of Physics, Hong Kong University of Science and Technology, Hong Kong, China}
\author{Ding Pan}
\email{dingpan@ust.hk}
\affiliation{Department of Physics, Hong Kong University of Science and Technology, Hong Kong, China}
\affiliation{Department of Chemistry, Hong Kong University of Science and Technology, Hong Kong, China}
\affiliation{HKUST Fok Ying Tung Research Institute, Guangzhou, China}

\date{\today}

\begin{abstract}
The chemistry of carbon in aqueous fluids at extreme pressure and temperature conditions is of great importance to Earth's deep carbon cycle, which substantially affects the carbon budget at Earth's surface and global climate change. At ambient conditions, the concentration of carbonic acid in water is negligible, so aqueous carbonic acid was simply ignored in previous geochemical models. However, by applying extensive ab initio molecular dynamics simulations at pressure and temperature conditions similar to those in Earth's upper mantle, we found that carbonic acid can be the most abundant carbon species in aqueous CO$_2$ solutions at $\sim$10 GPa and 1000 K. The mole percent of carbonic acid in total dissolved carbon species increases with increasing pressure along an isotherm, while its mole percent decreases with increasing temperature along an isobar. In CO$_2$-rich solutions, we found significant proton transfer between carbonic acid molecules and bicarbonate ions, which may enhance the conductivity of the solutions. The effects of pH buffering by carbonic acid may play an important role in water-rock interactions in Earth's interior. Our findings suggest that carbonic acid is an important carbon carrier in the deep carbon cycle. 

\end{abstract}

\maketitle

\section{Introduction}
The global carbon cycle is of great importance to Earth's climate, human's energy consumption, and sustainable development. Although a tremendous number of studies have focused on the carbon cycle in the atmosphere, oceans, and the shallow crust, less is known about the carbon cycle in Earth's deep interior \cite{hazen2013deep}, which may host more than 90\% of Earth's carbon \cite{Falkowski2000} and also actively exchanges carbon with Earth's near-surface reservoirs through volcanism and subduction \cite{ Manning2014, Kelemen2015}.
In the deep carbon cycle, fluids containing water as a primary component play a key role in transporting carbon in the deep mantle and back to magmas in volcanoes and eventually to the atmosphere as carbon dioxide (CO$_2$) \cite{manning2004chemistry,Manning2014}.
However, many basic questions about carbon-bearing aqueous fluids are poorly known; for example, what are the aqueous carbon species in the deep carbon cycle? A major obstacle to understanding deep carbon transport is the lack of knowledge of carbon reactions in water at the extreme conditions found in Earth's deep interior \cite{manning2013chemistry}.

Both pressure (P) and temperature (T) increase with increasing depth inside Earth. For example, in the upper mantle, pressure can reach $\sim$13 GPa and temperature $\sim$1700 K \cite{Frost2008}. Most crust and mantle fluids are at vapor-liquid supercritical conditions, and their properties differ fundamentally from those at ambient or near-ambient conditions, which has been shown by many recent studies \cite{manning2018}. For oxidized carbon dissolved in water, many geochemical models assume that CO$_2$(aq) is the major carbon species, e.g., Refs \cite{Mader1991, Duan2006, Zhang2009, Holland2011}, but recent studies show that carbon-containing ions, HCO$_3^-$ and CO$_3^{2-}$, may be more abundant than CO$_2$(aq) in Earth's mantle \cite{Facq2014, Pan2016}.

Another possible oxidized carbon species is carbonic acid (H$_2$CO$_3$) \cite{adamczyk2009real, loerting2010aqueous}. At ambient conditions, dry H$_2$CO$_3$ is very stable with a half-life as long as 180,000 years \cite{Loerting2000}, whereas in water it decomposes rapidly and its concentration is only about 0.1\% of CO$_2$(aq) concentration \cite{manning2013chemistry}; thus, in geochemical modeling H$_2$CO$_3$(aq) has simply been treated as CO$_2$(aq), except in a very recent model \cite{huang2019extended}. At 2.4 GPa and 553 K, Wang et al. found possible spectroscopic evidence of H$_2$CO$_3$(aq), suggesting it might be an important species in aqueous CO$_2$ solutions under extreme conditions \cite{Wang2016}, whereas Abramson et al. commented that the new vibrational band could be also attributed to HCO$_3^-$ at high concentration \cite{Abramson2017a}. Note that these experimental P-T conditions can only be found in the shallow or cold mantle areas. It is very challenging to experimentally detect carbon species at \emph{both} high pressures (HP) \emph{and} high temperatures (HT), as found in the deep mantle. In our previous theoretical work, we applied ab initio molecular dynamics simulations to study CO$_2$ in water at $\sim$11 GPa and 1000 K, and found that about 80\% of dissolved carbon is HCO$_3^-$ and 10$\sim$20\% H$_2$CO$_3$(aq) as a minor product \cite{Pan2016}. The solution is very dilute with x(CO$_2$) = 0.016. In fact, carbon-bearing fluids in deep Earth may range from water-rich to CO$_2$-rich \cite{andersen2001fluid}. The carbon speciation in CO$_2$-rich aqueous fluids under Earth's mantle conditions is still unknown.

Here, by applying extensive ab initio molecular dynamics (AIMD) simulations, we studied the CO$_2$-H$_2$O mixtures at various concentrations from 3 to 11 GPa and between 1000 and 1400 K, P-T conditions similar to those in subduction zones in Earth's upper mantle. Our total simulation time exceeds 3.8 nanoseconds. We found that at $\sim$10 GPa and 1000 K, H$_2$CO$_3$(aq) can be the predominant carbon species in the aqueous CO$_2$ solutions. With increasing pressure along an isotherm, the mole percent of H$_2$CO$_3$(aq) in total dissolved carbon species increases, while with increasing temperature along an isobar, its mole percent decreases. We show the possible P-T range where H$_2$CO$_3$(aq) may exist. The rich chemistry of H$_2$CO$_3$(aq) may play an important role in the dissolution of carbonate minerals in aqueous fluids in deep Earth. Our findings suggest that H$_2$CO$_3$(aq) is an important carbon carrier in the deep carbon cycle. 

\section{Results and discussion}
\subsection{Equation of state of CO$_2$-H$_2$O mixtures}
First, we calculated the equation of state (EOS) of CO$_2$-H$_2$O mixtures given by density functional theory (DFT) with the PBE exchange-correlation functional. In previous studies we used the same level of theory to calculate many properties of water under HP-HT conditions, such as equation of state \cite{Pan2013}, dielectric constant \cite{Pan2014}, Raman and IR spectra, and ionic conductivity \cite{Rozsa2018}, as well as the reactions of CO$_2$ and carbonates in HP-HT water \cite{Pan2016}. In Table SI, we compare the DFT pressures of CO$_2$-H$_2$O systems with those obtained by popular geochemical models \cite{Duan2006, Zhang2009} and classical force-field MD simulations at various CO$_2$ concentrations. The force field is from Duan and Zhang's work in 2006 \cite{Duan2006}. Since the EOS in Ref. \cite{Duan2006} was obtained by fitting the MD simulation data using the force field and the experimental data, it is not surprising the pressures calculated by Duan and Zhang's EOS in Ref. \cite{Duan2006} are in excellent agreement with those calculated by our MD simulations with the same force field. The EOS pressures of Ref. \cite{Duan2006} are generally smaller than the results obtained by Zhang and Duan's EOS \cite{Zhang2009} in 2009, which was designed for C-O-H fluids in Earth's mantle. It was recently reported \cite{Zhang2016} that for CO$_2$-H$_2$O binary systems, the EOS in Ref. \cite{Duan2006} performs better than the one in Ref. \cite{Zhang2009}.

In our simulations, we varied the initial mole fraction of CO$_2$(aq), which is defined as $\mathrm{x(CO_2)} = \frac{n\mathrm{(CO_2)}}{n\mathrm{(CO_2)}+m\mathrm{(H_2O)}}$, where $n\mathrm{(CO_2)}$ and $m\mathrm{(H_2O)}$ are the numbers of CO$_2$ and water molecules, respectively. When x(CO$_2$) is smaller than 0.1, our DFT pressures are slightly larger than those obtained by the EOS in Ref. \cite{Duan2006}, while for x(CO$_2$) larger than 0.1, the DFT pressures become smaller. Note that in the simulations using the force field from Ref. \cite{Duan2006}, no chemical reaction occurs between CO$_2$ and water, whereas in the ab initio simulations the carbon atoms may be converted from sp to sp$^2$ hybridization, and vice versa. The newly generated carbon species may have different molecular volumes, which affect the EOS of the mixtures. Overall, the agreement between our ab initio data and available EOS data is satisfactory, indicating that the PBE functional is not only suitable for water under extreme conditions, but may also be used to study CO$_2$-H$_2$O mixtures at HP and HT.

\subsection{Speciation of CO$_2$-H$_2$O mixtures}
\begin{figure}
\centering
\vspace{5mm}
\includegraphics[width=0.5\textwidth]{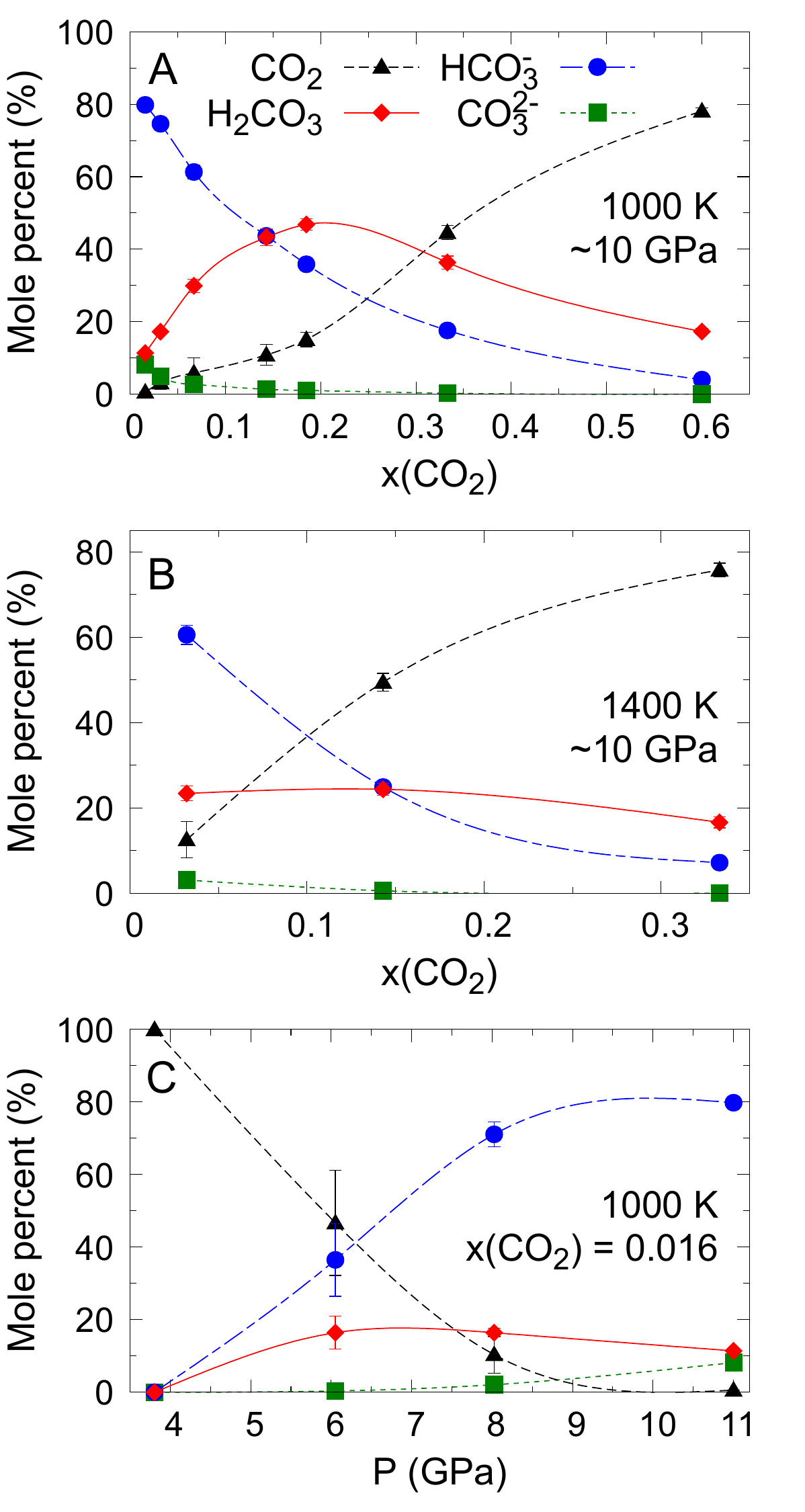}
\caption{Mole percents of CO$_2$(aq), H$_2$CO$_3$(aq), HCO$_3^-$, and CO$_3^{2-}$ in total dissolved carbon species in water. (A) The mole percents of carbon species as functions of the initial mole fraction of CO$_2$(aq) (x(CO$_2$)) at $\sim$10 GPa and 1000 K. (B) The mole percents of carbon species as functions of x(CO$_2$) at $\sim$10 GPa and 1400 K.
(C) The mole percents of carbon species as functions of pressure at 1000 K and x(CO$_2$) = 0.016. Simulation data points are interpolated by cubic splines. 
Error bars are obtained by using the blocking method \cite{Flyvbjerg1989}. 
} 
\label{carbon}
\end{figure}

Fig. \ref{carbon}(A) shows the relative concentrations of carbon species as functions of x(CO$_2$) obtained from AIMD simulations at  $\sim$10 GPa and 1000 K. The mole percents of CO$_2$(aq), HCO$_3^-$, CO$_3^2-$, and H$_2$CO$_3$(aq) as functions of simulation time are shown in Fig. S1. With increasing x(CO$_2$), the mole percent of HCO$_3^-$ in total dissolved carbon species decreases, and more CO$_2$(aq) molecules were found in the simulations. The crossover between  HCO$_3^-$ and CO$_2$(aq) happens when x(CO$_2$) is about 0.25. We found few carbonate ions, and the mole percent of CO$_3^{2-}$ is less than 10\% in all the simulations in Fig. \ref{carbon}. Interestingly, when x(CO$_2$) is in the range of 0.14 to 0.31, carbonic acid becomes the predominant carbon species, more abundant than any other carbon-containing molecules or ions. When increasing x(CO$_2$) up to $\sim$0.20, the mole percent of H$_2$CO$_3$(aq) reaches its maximum ($\sim$47\%); at higher x(CO$_2$), the mole percent of H$_2$CO$_3$ slowly decreases. During the decrease, H$_2$CO$_3$(aq) is always more abundant than HCO$_3^-$. After increasing the temperature from 1000 K to 1400 K at $\sim$10 GPa, the HCO$_3^-$-CO$_2$ crossover happens at a lower carbon concentration with x(CO$_2$) = 0.10 (Fig. \ref{carbon}(B)). 
At 1400 K, the mole percent of H$_2$CO$_3$(aq) is about 20\% when x(CO$_2$) is between 0.03 and 0.33, and does not have a clear maximum value.

\subsection{Molecular geometry of H$_2$CO$_3$}
\begin{figure}
\centering
\includegraphics[width=0.7\textwidth]{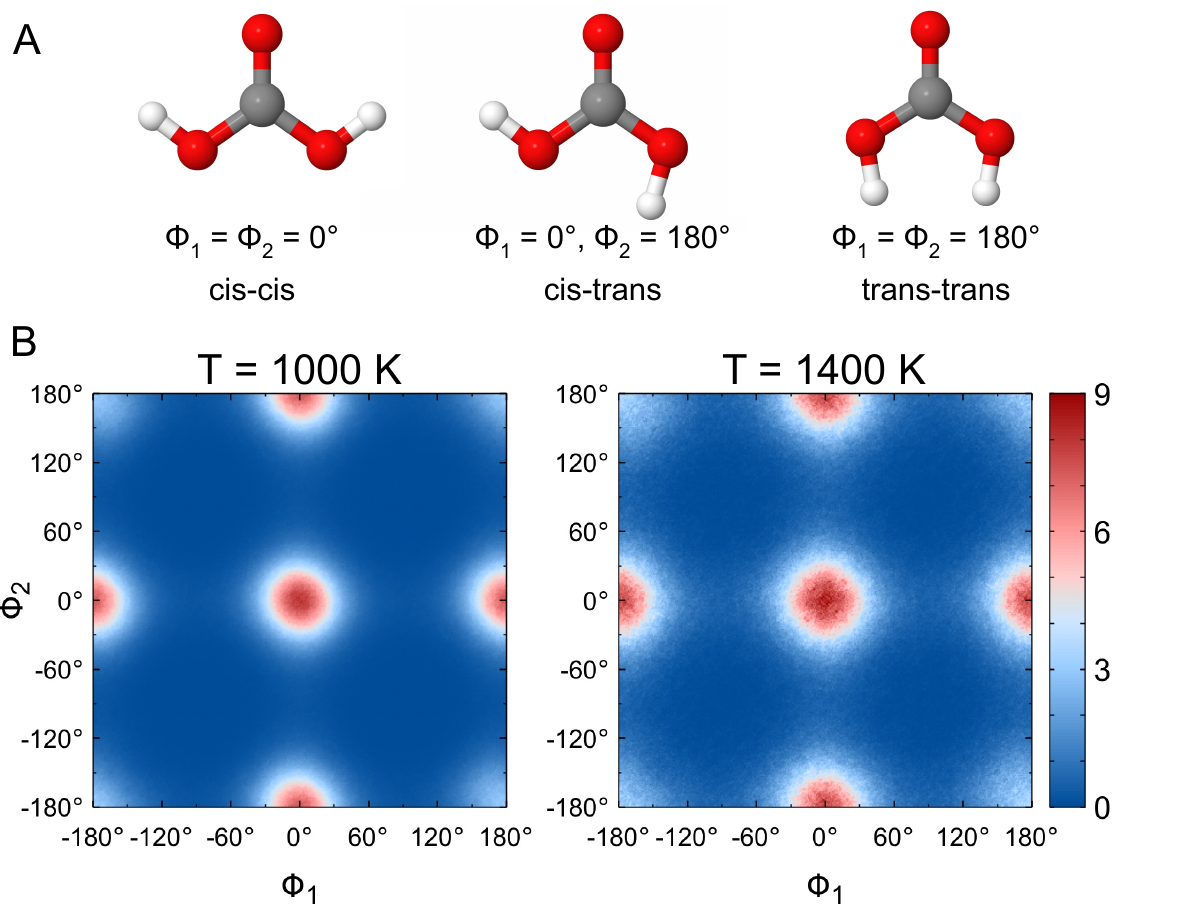}
\caption{ Molecular geometry of H$_2$CO$_3$ dissolved in water at $\sim$10 GPa, 1000 and 1400 K. (A) Three H$_2$CO$_3$ conformers. The dihedral angles, $\phi_1$ and $\phi_2$, are between the O=C and O-H bonds. 
(B) Probability densities of $\phi_1$ and $\phi_2$ (unit: 10$^{-5}$/degree$^{2}$).    
}
\label{dihedral}
\end{figure}

The carbonic acid molecule has one carbon atom and three oxygen atoms, which form the trigonal planar shape \cite{Reisenauer2014, bucher2014clarifying}. Two hydrogen atoms, which are each bonded to an oxygen atom, appear not always in the plane of the carbon and oxygen atoms in the MD simulations. Fig. \ref{dihedral} shows the distribution of the two O=C-O-H dihedral angles. The H$_2$CO$_3$ conformer with the lowest free energy is the cis-cis conformer, where the two dihedral angles are both 0$^\circ$. The conformer with the second lowest free energy is cis-trans, where the two dihedral angles are 0$^\circ$ and 180$^\circ$. The trans-trans conformer, where the two dihedral angles are both 180$^\circ$, has much higher free energy than the former two. We calculated the free energy difference between the $i$th and $j$th conformer by $\Delta F = -k_BT \ln \left( \frac{P_i}{P_j} \right)$, where $k_B$ is the Boltzmann constant and $P_i$ is the probability of the $i$th conformer. The free energies are calculated with respect to the cis-cis conformer. At $\sim$10 GPa and 1000 K, the free energy of the cis-trans conformer is 0.14 kcal/mol (0.006 eV), while the trans-trans conformer has the free energy of 1.82 kcal/mol (0.079 eV). When increasing the temperature from 1000 to 1400 K along the isobar of $\sim$10 GPa, the free energy of the cis-trans conformer increases to 0.19 kcal/mol (0.008 eV), while the trans-trans one increases to 2.31 kcal/mol (0.100 eV).

In the gas phase at 0 K, the free energies of the cis-trans and trans-trans conformers are 1.6 and 10.1 kcal/mol respectively \cite{Reisenauer2014}, much higher than the respective free energies in the aqueous solutions studied here. This may be attributed to the dielectric screening of water. The three H$_2$CO$_3$ conformers have different electrostatic interactions.  The two hydrogen atoms with partial positive charges in the trans-trans conformer are closer than in the other two conformers, so the stronger electrostatic repulsive force causes a higher energy in the trans-trans conformer.
Because the dielectric screening of water weakens electrostatic interactions, we see the lower free energies for the trans-trans and cis-trans conformers in water than in the gas phase. When increasing T along an isobar, the dielectric constant of water decreases \cite{Pan2013}, so the screening becomes weaker and the free energies become larger again.

\subsection{Reaction mechanisms of H$_2$CO$_3$(aq)}
\begin{figure}
\centering
\vspace{5mm}
\includegraphics[width=0.7\textwidth]{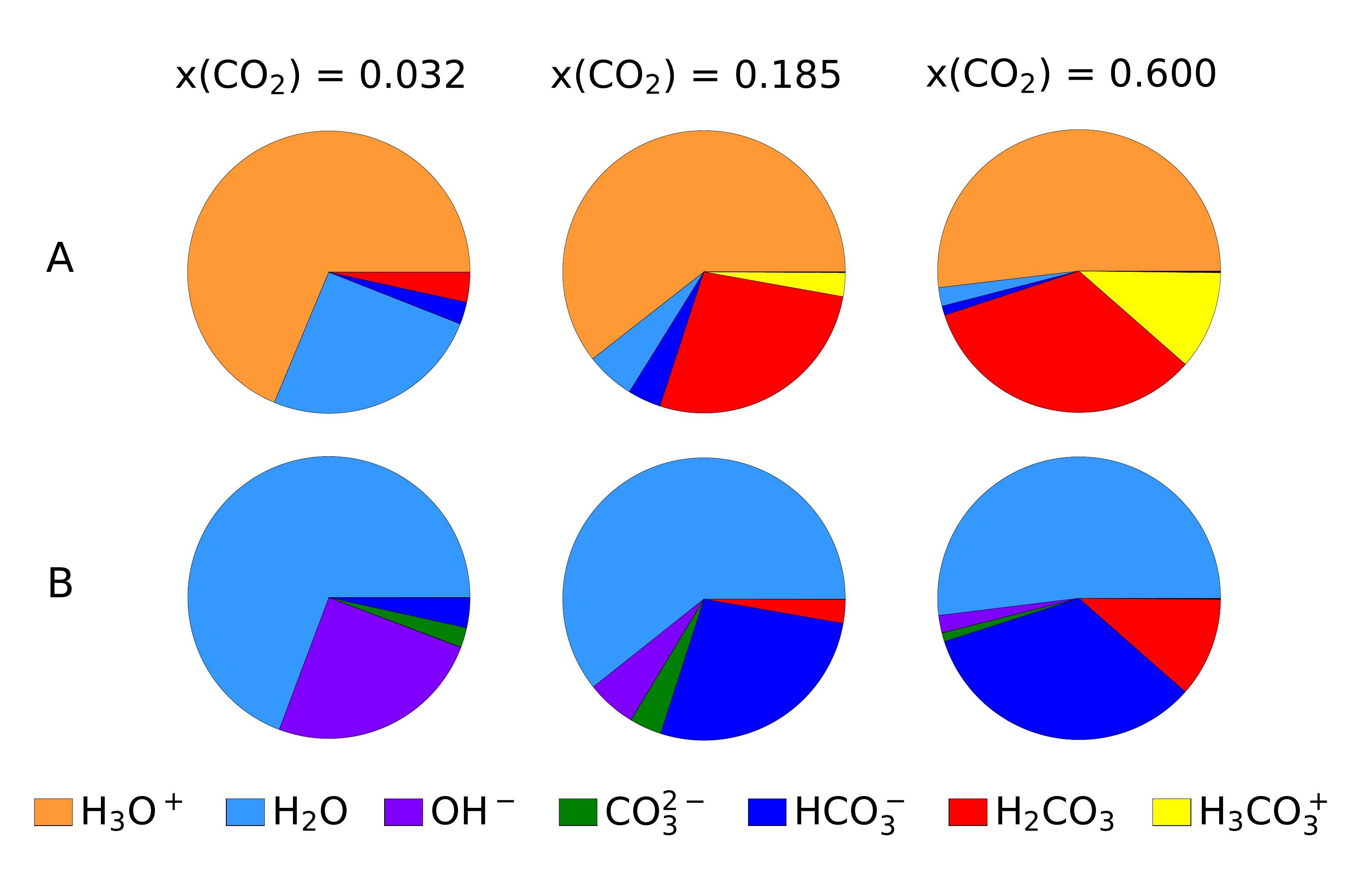}
\caption{Compositions of proton-donating and proton-accepting molecules or ions in H$_2$CO$_3$(aq) formation and dissociation reactions, respectively, at $\sim$10 GPa and 1000 K. (A) Molecules or ions which donate protons to the reaction: $\mathrm{HCO_3^- + H^+ \rightarrow H_2CO_3(aq) }$. (B) Molecules or ions which accept protons from the backward reaction in (A). 
}
\label{formation_proton}
\end{figure}

In Fig. \ref{formation_proton}, we analyzed how H$_2$CO$_3$(aq) forms and dissociates at $\sim$10 GPa and 1000 K. In our AIMD simulations, $\sim$90\% of the H$_2$CO$_3$(aq) formation reactions follow a two-step mechanism \cite{Stirling2010}: (i) the reaction of CO$_2$(aq) with OH$^-$ or water to form HCO$_3^-$, 
\begin{align}
          \mathrm{CO_2(aq)+OH^-} &\rightleftharpoons \mathrm{HCO_3^-} \label{bicarOH}\\
          \mathrm{CO_2(aq)+H_2O} &\rightleftharpoons \mathrm{HCO_3^- + H^+} \label{bicarH2O}
\end{align}
and (ii) the protonation of HCO$_3^-$ to form H$_2$CO$_3$(aq):
\begin{equation}
  \mathrm{HCO_3^- + H^+ \rightleftharpoons H_2CO_3(aq) }
\label{carbonic}
\end{equation}
Because the solutions are acidic, it is very rare to see free OH$^-$. In reaction (\ref{bicarOH}), the OH$^-$ ion is often associated with the H$_3$O$^+$ ion generated by the self-ionization of water. If the proton keeps hopping between the OH$^-$ and H$_3$O$^+$ ions, reaction (\ref{bicarOH}) will also go backward and forward frequently. When the proton in H$_3$O$^+$ transfers away from the ion pair along a hydrogen-bond wire via the Grotthuss mechanism \cite{Marx2006}, the chemical equilibrium of reaction (\ref{bicarOH}) shifts to the right. Similarly, in reaction (\ref{bicarH2O}) where the CO$_2$ molecule reacts with a water molecule, if the released proton is accepted by a third species and cannot transfer back, the generated HCO$_3^-$ will not often change back to CO$_2$(aq).

\begin{figure}
\centering
\vspace{5mm}
\includegraphics[width=0.7\textwidth]{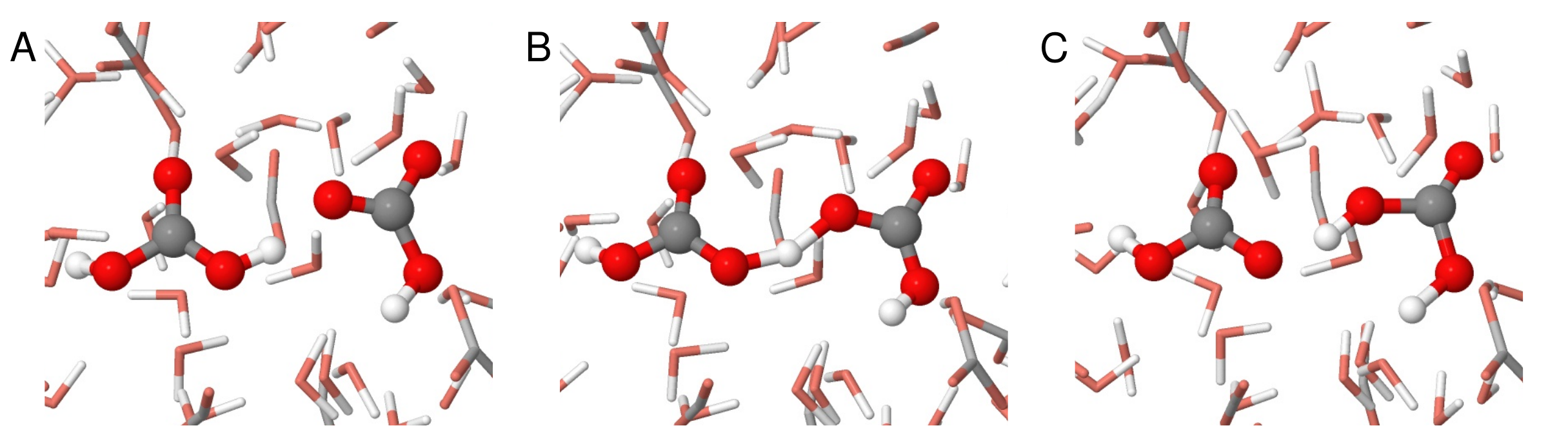}
\caption{Snapshots of the proton transfer from H$_2$CO$_3$(aq) to HCO$_3^-$. (A) A cis-cis H$_2$CO$_3$ molecule is approaching a HCO$_3^-$ ion. (B) A proton is hopping from H$_2$CO$_3$(aq) to HCO$_3^-$. (C) HCO$_3^-$ and trans-trans H$_2$CO$_3$ are formed after the reaction.}
\label{proton_transfer}
\end{figure}

In reaction (\ref{carbonic}), the HCO$_3^-$ ion accepts one proton to become H$_2$CO$_3$(aq). Fig. \ref{formation_proton}(A) shows that when the carbon concentration is low, most of the protons are from H$_3$O$^+$, whereas with increasing x(CO$_2$), more and more protons come from H$_2$CO$_3$(aq), indicating that protons can transfer between HCO$_3^-$ and H$_2$CO$_3$(aq) in CO$_2$-rich solutions (see Fig. \ref{proton_transfer}). We even found H$_3$CO$_3^+$ as an intermediate proton donor with lifetime less than 0.15 ps. In fact, about 10\% of H$_2$CO$_3$(aq) is generated from the dissociation of H$_3$CO$_3^+$:
\begin{equation}
  \mathrm{H_3CO_3^+  \rightleftharpoons H_2CO_3(aq)  + H^+}
\label{H3CO3}
\end{equation}
The sp$^2$ carbon species such as H$_2$CO$_3$, HCO$_3^-$, and H$_3$CO$_3^+$ may form a network to conduct protons in water. It was expected that the ionic conductivity of CO$_2$-H$_2$O mixtures decreases significantly with increasing x(CO$_2$), if only the self-ionization of water contributes to ionic conduction \cite{manning2018}. This expectation may not be valid, as the generated sp$^2$ carbon species can enhance the ionic conductivity.

At $\sim$10 GPa and 1000 K, about 90\% of H$_2$CO$_3$(aq) dissociates to become HCO$_3^-$ via backward reaction (\ref{carbonic}). The lifetime of H$_2$CO$_3$(aq) is typically less than 1 ps (see Fig. S4). When the solutions are dilute, most of protons transfer to water molecules to form H$_3$O$^+$, whereas with increasing x(CO$_2$), the role of HCO$_3^-$ or H$_2$CO$_3$(aq) as proton acceptors becomes more important, as shown in Fig. \ref{formation_proton}(B). Interestingly, only 1$\sim$3\% of HCO$_3^-$ further decompose into CO$_2$(aq) (see Fig. S5). We found two decomposition pathways: backward reaction (\ref{bicarOH}) and (\ref{bicarH2O}). Decomposition via backward reaction (\ref{bicarOH}) is more likely to happen when x(CO$_2$) is low. With increasing x(CO$_2$), decomposition via backward reaction (\ref{bicarH2O}) becomes more frequent. In backward reaction (\ref{bicarOH}), if the OH$^-$ ion remains in the first solvation shell of CO$_2$, the O$_2$C-OH bond often keeps breaking and forming unless the OH$^-$ ion accepts a proton and becomes a H$_2$O molecule. Averagely, CO$_2$(aq) generated through backward reaction (\ref{bicarH2O}) has a longer lifetime than that from backward reaction (\ref{bicarOH}). About 10\% of H$_2$CO$_3$(aq) does not dissociate or decompose, but receives protons to become the intermediate H$_3$CO$_3^+$ through backward reaction (\ref{H3CO3}). In the gas phase, the decomposition of H$_2$CO$_3$ occurs via the concerted mechanism that involves the dehydroxylation of one OH$^-$ group and the deprotonation of the other OH$^-$ group simultaneously \cite{Loerting2000, Kumar2007}. Here, we found only the stepwise process through backward reactions (\ref{carbonic}), and (\ref{bicarH2O}) or (\ref{bicarOH}). Galib and Hanna studied the decomposition mechanism of H$_2$CO$_3$(aq) using water clusters with varied sizes at ambient conditions, and found that the hydrogen bonding environment around the H$_3$O$^+$ ion as found in bulk water leads to stepwise decomposition \cite{Galib2014}, which indicates that at the supercritical conditions studied here there are still strong water-ion bonding interactions.

The reaction mechanisms of H$_2$CO$_3$(aq) formation and dissociation help us understand why H$_2$CO$_3$(aq) can be the predominant carbon species in supercritical water at $\sim$10 GPa and 1000 K. When the CO$_2$-H$_2$O solution is dilute, most CO$_2$ reacts and becomes HCO$_3^-$ via reaction (\ref{bicarOH}) or (\ref{bicarH2O}) \cite{Pan2016}. With increasing x(CO$_2$), the solution becomes more acidic and HCO$_3^-$ stays closer to H$_3$O$^+$, so reaction (\ref{carbonic}) is more likely to happen. Thus, when increasing x(CO$_2$) up to $\sim$0.20, the concentration of H$_2$CO$_3$(aq) increases to its maximum. However, when further increasing x(CO$_2$), a significant amount of CO$_2$(aq) does not react and the mole percent of HCO$_3^-$ in total dissolved carbon species keeps decreasing. As a result, the mole percent of H$_2$CO$_3$(aq) also decreases. When x(CO$_2$) $>$ 0.14, H$_2$CO$_3$(aq) is more abundant than HCO$_3^-$, indicating that H$_2$CO$_3$(aq) is energetically favored over HCO$_3^-$ in CO$_2$-rich solutions under extreme conditions.

\subsection{P-T range of H$_2$CO$_3$(aq)}
When increasing the temperature from 1000 K to 1400 K at $\sim$10 GPa, the mole percent of H$_2$CO$_3$(aq) becomes smaller. The concentration of CO$_2$(aq) at thermal equilibrium at 1400 K is higher than that at 1000 K when the initial CO$_2$ concentration is the same. As a result, the H$_2$CO$_3$ formation reactions are less likely to happen. Big molecules or ions tend to decompose at high temperatures, so at 1400 K, 9$\sim$10\% of HCO$_3^-$ decomposes into CO$_2$(aq), almost one order of magnitude more than that at 1000 K. 

Not only temperature but also pressure affects the presence of H$_2$CO$_3$(aq). We lowered the pressure from $\sim$10 GPa to $\sim$8.0 GPa, $\sim$6.1 GPa, and $\sim$3.8 GPa in the AIMD simulations at 1000 K and x(CO$_2$) = 0.016 (see Fig. \ref{carbon}(C)). Because the solutions are dilute, H$_2$CO$_3$(aq) is not the predominant carbon species, but we can estimate the formation of H$_2$CO$_3$(aq) with increasing x(CO$_2$) based on the reaction mechanisms discussed above. At $\sim$8.0 GPa, most of the CO$_2$ molecules still convert into HCO$_3^-$, but the concentration of HCO$_3^-$ is lower than that at $\sim$10 GPa; thus, with increasing x(CO$_2$), we also expect to have a significant amount of H$_2$CO$_3$(aq), despite a lower concentration than that at $\sim$10 GPa. Fig. \ref{carbon}(C) shows that when the pressure is lower than $\sim$6.3 GPa, CO$_2$(aq) becomes predominant. Because CO$_2$(aq) reacts slowly at low pressures, the transition pressure of $\sim$6.3 GPa should be an upper limit. Generally, with decreasing pressure at 1000 K, the mole percent of H$_2$CO$_3$(aq) decreases.

\begin{figure}
\centering
\vspace{5mm}
\includegraphics[width=0.8\textwidth]{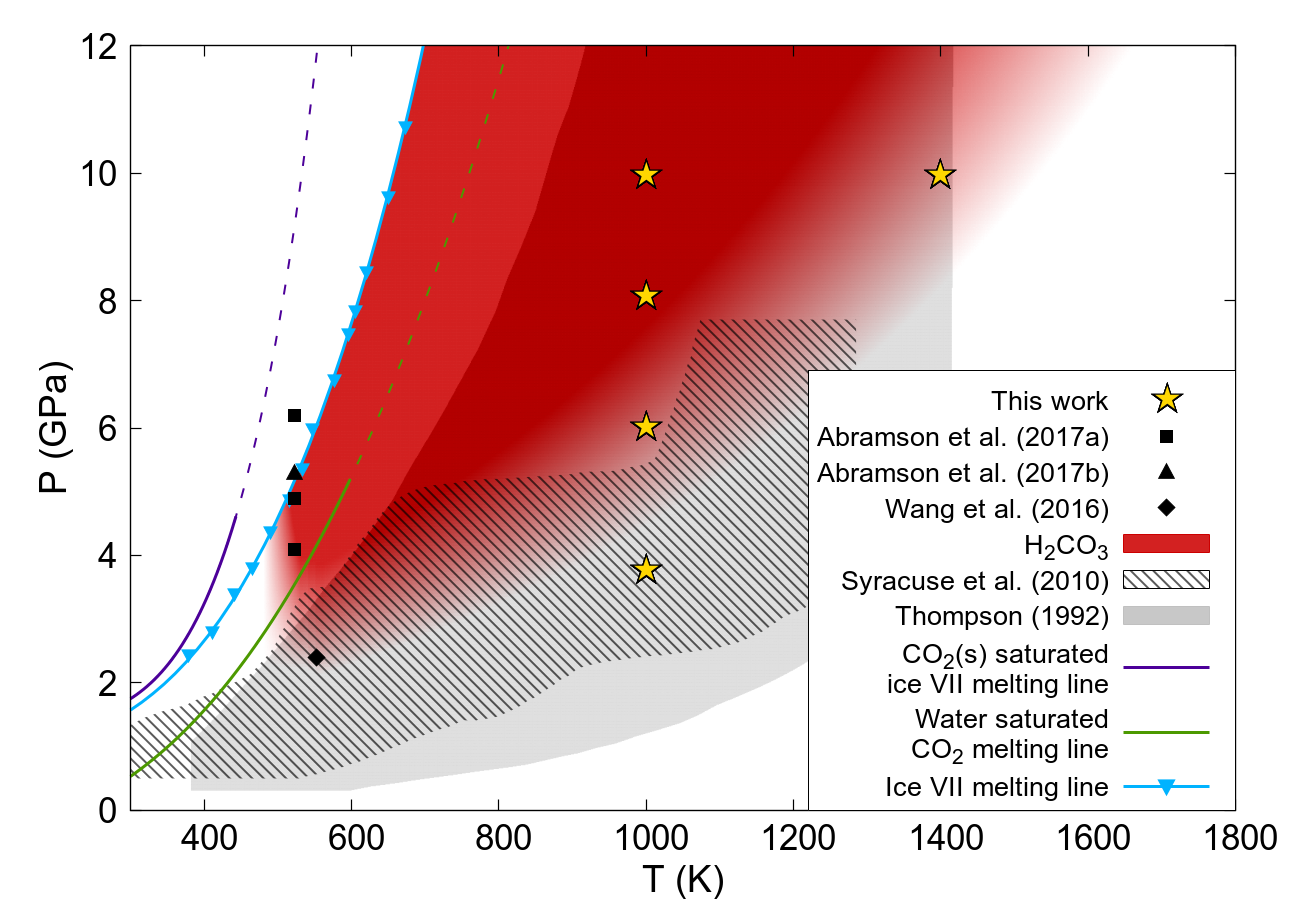}
\caption{Pressure-temperature range (in red) where H$_2$CO$_3$(aq) is an important solute in water. The P-T conditions in the possible experimental observations of H$_2$CO$_3$ are shown: Wang et al. \cite{Wang2016}, Abramson et al. (2017a) \cite{Abramson2017a}, and Abramson et al. (2017b) \cite{Abramson2017b}. The grey shaded and hatched areas show the P-T conditions of water in Earth's upper mantle \cite{Thompson1992} and subducting slab surfaces (W1300 model) \cite{Syracuse2010}, respectively. The ice VII melting line is from Ref. \cite{Dubrovinskaia2003}. The CO$_2$-saturated water and water-saturated CO$_2$ melting lines are from Ref. \cite{Abramson2017}. Dashed lines are extrapolation.} 
\label{phase}
\end{figure}

In Fig. \ref{phase}, we show the possible P-T range for H$_2$CO$_3$(aq) as a significant solute in water based on our AIMD simulations and experimental findings. With increasing P along an isotherm, the concentration of H$_2$CO$_3$(aq) increases, while with increasing T along an isobar, less H$_2$CO$_3$(aq) forms. There can be a significant amount of H$_2$CO$_3$(aq) at P $\sim$10 GPa and T $\leq$ 1400 K. Wang et al.  showed possible spectroscopic evidence of H$_2$CO$_3$(aq) at 2.4 GPa and 553 K in a mixture of CO$_2$ and H$_2$O \cite{Wang2016}. For aqueous CO$_2$ solutions, Abramson et al. found that with increasing P from 4.1 GPa to 6.2 GPa at 523 K, the Raman signal of CO$_2$(aq) gradually disappears and a new peak at $\sim$1040 cm$^{-1}$ might be attributed to high concentration HCO$_3^{-}$ or H$_2$CO$_3$(aq) \cite{Abramson2017a}. No evidence for the formation of H$_2$CO$_3$(aq) below 2.4 GPa is reported experimentally, so in Fig. \ref{phase} we use 2.4 GPa and 523 K as the P-T boundary of H$_2$CO$_3$(aq). In Fig. \ref{phase}, we also show the P-T conditions of water in Earth's upper mantle \cite{Thompson1992} and subducting slab surfaces \cite{Syracuse2010}, which largely overlap with the P-T range of the formation of H$_2$CO$_3$(aq). This indicates that there can be a significant amount of H$_2$CO$_3$(aq) in aqueous geofluids in subduction zones in the upper mantle. Dissolved CO$_2$ in aqueous geofluids may be gradually converted into H$_2$CO$_3$(aq) and HCO$_3^-$/CO$_3^{2-}$ with increasing depth. The sp$^2$-carbon-bearing fluids may also be brought back to Earth's surface by volcanism and the CO$_2$ degassing would occur in shallow areas.

In the DFT calculations, we used the semilocal exchange-correlation functional PBE \cite{Perdew1996}. It has been reported that generalized gradient approximations (GGAs) may not be able to describe doubly charged anions in water accurately at ambient conditions \cite{wan2014electronic}. However, PBE performs better at HP-HT conditions than at ambient conditions, as shown in many of our previous studies \cite{Pan2013, Pan2014, Pan2016, Rozsa2018}. Our previous finding that CO$_2$(aq) is absent at $\sim$11 GPa and 1000 K \cite{Pan2016} was further extended to lower pressures and temperatures by experiment \cite{Abramson2017a}, indicating that our computational method is capable to predict the properties of aqueous carbon solutions under extreme conditions. To validate our results, we compared the PBE simulations with the simulations using the hybrid functional PBE0 \cite{adamo1999toward,Pan2016}, which often gives results in better agreement with experimental ones due to the smaller charge delocalization error \cite{Cohen2008}. When we dissolved CO$_2$ in water (x(CO$_2$) = 0.016) at $\sim$11 GPa and 1000 K , we found that in both PBE and PBE0 simulations, HCO$_3^-$ is the predominant carbon species and its mole percents are very similar: 79.8\% vs. 75.0\% \cite{Pan2016}. PBE0 predicts even more H$_2$CO$_3$(aq) than PBE. Thus, our finding that H$_2$CO$_3$(aq) can be the most abundant carbon species under extreme conditions would not be affected by the PBE error. 

In addition, van der Waals interactions are not well captured by the PBE and PBE0 functionals. However, previous studies  showed that PBE and van der Waals functionals give closer results at high pressure than at ambient pressure for the equilibrium volume and dielectric constant of ice \cite{murray2012dispersion}, indicating that for water and ice under pressure, the dispersion interactions are not so important as those at ambient conditions. Besides, the breaking and forming of covalent bonds are little affected by weak van der Waals interactions, so the lack of dispersion interactions should not change our main results here.

The chemical reactions involving H$_2$CO$_3$(aq) greatly affect the acidity of aqueous carbon solutions, which may have an important impact on water-rock interactions in deep Earth \cite{manning2013thermodynamic}. It has been estimated that about half of carbon in subducting slabs or even more is brought back to Earth's surface \cite{Kelemen2015}. The previous geochemical studies on the devolatilization in subduction zones considered only CO$_2$(aq) as the major carbon species in water \cite{Manning2014}. If so, the amounts of carbon removed from subducting slabs are too small to account for the large volumes of carbon found in magmas \cite{Manning2014}. It was recently reported that significant quantities of carbonate minerals are dissolved by the infiltrating fluids, and are brought back to magmas, which may potentially balance CO$_2$ emissions from volcanoes \cite{Ague2014}. However, the chemical mechanism of this dissolution process is not yet known. At HP-HT conditions, the solubilities of carbonate minerals in pure water increase, but are still not very large \cite{Facq2014,Pan2013}. If we consider only molecular CO$_2$ in the solutions, the solubilities should become even lower \cite{manning2013chemistry}. Our current work suggests that the effects of pH buffering by H$_2$CO$_3$(aq) may play an important role in enhancing the dissolution of carbonate minerals \cite{manning2013thermodynamic}.

\section{Conclusion}
In conclusion, we conducted ab initio molecular dynamics simulations to study dissolved CO$_2$ in both water-rich and CO$_2$-rich aqueous solutions from 3 to 11 GPa and at 1000 and 1400 K, P-T conditions similar to those found in subduction zones in Earth's upper mantle. We found that H$_2$CO$_3$(aq), whose concentration is negligible in water at ambient conditions, is the most abundant carbon species in the aqueous CO$_2$ solutions at $\sim$10 GPa and 1000 K, when the initial mole fraction of CO$_2$(aq) is between 0.14 and 0.31. With increasing pressure along an isotherm, the mole percent of H$_2$CO$_3$(aq) in total dissolved carbons species increases, while with increasing temperature along an isobar, its mole percent decreases. Contrary to popular geochemical models assuming carbon mainly exists in the form of molecular CO$_2$ in aqueous geofluids, our study suggests that big carbon-containing molecules like H$_2$CO$_3$(aq) cannot be ignored under deep Earth conditions. We found significant proton transfer between H$_2$CO$_3$(aq) and HCO$_3^-$ in CO$_2$-rich solutions, which may enhance the ionic conductivity of the fluids. The rich chemistry of H$_2$CO$_3$(aq) may substantially affect water-rock interactions in deep Earth. For example, the effects of pH buffering by H$_2$CO$_3$(aq) may play an important role in enhancing the dissolution of carbonate minerals. Our findings suggest that H$_2$CO$_3$(aq) is an important carbon carrier in the deep carbon cycle. Not only inside Earth, it may also largely exist in other water-rich planets \cite{Marounina2019}. 

\section{Methods}
\subsection{Ab initio molecular dynamics (AIMD)}
AIMD simulations were carried out with the Qbox code  (version 1.63.8, http://qboxcode.org/) \cite{gygi2008} in the Born-Oppenheimer approximation. We used plane wave basis sets, the PBE exchange-correlation functional \cite{Perdew1996}, and norm-conserving pseudopotentials with a  kinetic energy cutoff of 85 Ry (Pseudopotential Table, http://fpmd.ucdavis.edu/potentials) \cite{Hamann1979, Vanderbilt1985}. In the pressure calculations, the cutoff was increased to 145 Ry. We sampled the Brillouin zone of the simulation box at the $\Gamma$ point only. We used deuterium instead of hydrogen to have a larger time step (0.24 fs) for computational convenience. The temperature was controlled by the Bussi-Donadio-Parrinello thermostat ($\tau$ = 24.2 fs) \cite{Bussi2007}. The numbers of CO$_2$ and water molecules in simulation boxes are listed in Table SI. In MD simulations, the equilibration time is 20 ps. Each production run is between 180 and 580 ps long, as shown in Fig. S1-S3. 

We used the atomic trajectories obtained from AIMD simulations to determine the chemical form of carbon species. To determine whether a carbon atom is sp or sp$^2$ hybridized, we sorted the three smallest C-O distances near each carbon atom. If the difference between the second and third C-O distances is more than 0.4 \AA, the species was defined as CO$_2$; otherwise, it was considered as a CO$_3^{2-}$ anion. For each hydrogen atom near carbon species, we looked for the closest oxygen atom to form the O-H bond. The bicarbonate ion and carbonic acid have one and two O-H bonds, respectively.

\subsection{Classical molecular dynamics}
Classical MD simulations were performed using simulation boxes containing 320$\sim$512 molecules. We used the Gromacs package version 5.1.2 \cite{Abraham2015}. The time step is 1 fs. The force fields for CO$_2$ and water are from Ref. \cite{Duan2006}. The temperature was maintained by the Bussi-Donadio-Parrinello thermostat ($\tau$ = 0.1 ps). Electrostatic interactions were calculated using the fast smooth particle-mesh Ewald method with a grid spacing of 1 {\AA} and cubic interpolation. A cutoff radius of 10 {\AA} was used for van der Waals interactions. Production runs are $\sim$2 ns after 1 ns equilibration.

\section{Acknowledgements}
We thank Giulia Galli and Viktor Rozsa for their helpful discussions. N.S. acknowledges the Hong Kong PhD Fellowship Scheme. D.P. acknowledges support from the Croucher Foundation through
the Croucher Innovation Award, Hong Kong Research Grants Council (Projects ECS-26305017, GRF-16307618), National Natural Science Foundation of China (Project 11774072), and the Alfred P. Sloan Foundation through the Deep Carbon Observatory.

\section{Contributions}
D.P. designed the research. N.S. performed the ab initio and classical simulations. All authors contributed to the analysis and discussion of the data and the writing of the manuscript.

\bibliography{references}

\newpage

\end{document}

\title{Supporting Information}

\author{Nore Stolte}
\affiliation{Department of Physics, Hong Kong University of Science and Technology, Hong Kong, China}
\author{Ding Pan}
\email{dingpan@ust.hk}
\affiliation{Department of Physics, Hong Kong University of Science and Technology, Hong Kong, China}
\affiliation{Department of Chemistry, Hong Kong University of Science and Technology, Hong Kong, China}
\affiliation{HKUST Fok Ying Tung Research Institute, Guangzhou, China}

\maketitle

\begin{table}[H]
\begin{minipage}{1.0\textwidth}
\caption{AIMD simulation conditions. The pressures are calculated from simulations and models. x(CO$_2$) is the initial mole fraction of CO$_2$(aq), and $n$(CO$_2$) and $m$(H$_2$O) are the numbers of CO$_2$ and H$_2$O molecules, respectively, in the simulation box. Numbers in parentheses are standard deviations.}
\label{pressurest}
\begin{tabular}{ c | c | c  c  c  c  c  c  c }
\hline
\hline
x(CO$_2$) & T (K) &  $n$(CO$_2$) & $m$(H$_2$O) & V (cm$^3$/mol) & P$^*$ (GPa) & P$^{\circ}$ (GPa) & P$^{\dagger}$ (GPa) & P$^{\ddagger}$ (GPa) \\
\hline
0.032 & 1000 & 2 & 60 & 11.59 & 10.26 & 12.32 & 10.5(0.2) & 10.7(1.0) \\
0.067 & 1000 & 4 & 56 & 11.89 & 10.26 & 12.46 & 10.6(0.2) & 10.3(1.1) \\
0.143 & 1000 & 8 & 48 & 12.59 & 10.26 & 12.49 & 10.7(0.2) & 9.7(1.2) \\
0.185 & 1000 & 10 & 44 & 12.99 & 10.26 & 12.39 & 10.7(0.2) & 9.4(1.2) \\
0.333 & 1000 & 16 & 32 & 14.48 & 10.26 & 11.71 & 10.6(0.2) & 9.3(1.2) \\
0.600 & 1000 & 24 & 16 & 17.24 & 10.26 & 10.31 & 10.6(0.2) & 10.0(1.2) \\
 & & & & & \\
0.032 & 1400 & 2 & 60 & 12.20 & 10.26 & 12.23 & 10.3(0.2) & 10.4(1.2) \\ 
0.143 & 1400 & 8 & 48 & 13.47 & 10.26 & 11.23 & 9.9(0.2) & 9.2(1.3) \\
0.333 & 1400 & 16 & 32 & 15.67 & 10.26 & 9.88 & 9.4(0.2) & 9.0(1.3) \\
 & & & & & \\
0.016 & 1000 & 1 & 63 & 15.22 & 3.00 & 3.28 & 3.0(0.1) & 3.8(0.8) \\
0.016 & 1000 & 1 & 63 & 13.50 & 5.00 & 5.65 & 5.0(0.2) & 6.1(0.9) \\
0.016 & 1000 & 1 & 63 & 12.49 & 7.00 & 8.10 & 7.0(0.2) & 8.0(1.0) \\
\hline 
\hline
\end{tabular}
\begin{flushleft}
$^*$ Duan and Zhang \cite{Duan2006}.\\
$^{\circ}$ Zhang and Duan \cite{Zhang2009}.\\
$^{\dagger}$ This work, force-field MD simulations.\\
$^{\ddagger}$ This work, AIMD simulations.
\end{flushleft}
\end{minipage}
\end{table}

\begin{table}
\centering
\begin{minipage}{1.0\textwidth}
\caption{Mole percents of carbon species in total dissolved carbon obtained from AIMD simulations (unit: \%). Numbers in parentheses are standard deviations.}
\label{concentrations}
\begin{tabular}{ c | c | c | c  c  c  c  c  c }
\hline
\hline
T (K) & P (GPa) & x(CO$_2$) & CO$_2$ & CO$_3^{2-}$ & HCO$_3^{-}$ & H$_2$CO$_3$ & H$_3$CO$_3^+$ & Pyrocarbonate \\
\hline
1000 & 10.7(1.0) & 0.032 & 3.2(1.8) & 4.9(0.4) & 74.6(1.6) & 17.3(0.8) & 0.0(0.0) & 0.0(0.0) \\
1000 & 10.3(1.1) & 0.067 & 5.9(4.1) & 2.7(0.1) & 61.3(1.9) & 29.9(1.9) & 0.1(0.0) & 0.0(0.0) \\
1000 & 9.7(1.2) & 0.143 & 10.9(2.8) & 1.4(0.1) & 43.7(0.8) & 43.3(2.3) & 0.5(0.1) & 0.3(0.1) \\
1000 & 9.4(1.2) & 0.185 & 15.1(2.0) & 1.1(0.0) & 35.9(0.7) & 46.8(1.5) & 0.8(0.1) & 0.3(0.1) \\
1000 & 9.3(1.2) & 0.333 & 44.6(2.0) & 0.3(0.0) & 17.6(0.3) & 36.4(1.8) & 0.8(0.1) & 0.3(0.1) \\
1000 & 10.0(1.2) & 0.600 & 78.1(0.9) & 0.0(0.0) & 4.1(0.2) & 17.3(0.8) & 0.4(0.1) & 0.1(0.0) \\
 & & & & & & & & \\
1400 & 10.4(1.2) & 0.032 & 12.6(4.3) & 3.1(0.3) & 60.6(2.3) & 23.4(1.8) & 0.1(0.0) & 0.3(0.3) \\
1400 & 9.2(1.3) & 0.143 & 49.5(2.1) & 0.6(0.0) & 24.9(0.7) & 24.4(1.4) & 0.3(0.1) & 0.4(0.2) \\
1400 & 9.0(1.3) & 0.333 & 75.8(1.6) & 0.0(0.0) & 7.2(0.3) & 16.6(1.3) & 0.2(0.0) & 0.1(0.0) \\
 & & & & & & & & \\
1000 & 3.8(0.8) & 0.016 & 100.0(0.0) & 0.0(0.0) & 0.0(0.0) & 0.0(0.0) & 0.0(0.0) & 0.0(0.0) \\
1000 & 6.1(0.9) & 0.016 & 46.7(14.5) & 0.4(0.1) & 36.5(10.0) & 16.4(4.5) & 0.0(0.0) & 0.0(0.0) \\
1000 & 8.0(1.0) & 0.016 & 10.4(5.1) & 2.1(0.2) & 71.1(3.4) & 16.4(1.2) & 0.0(0.0) & 0.0(0.0) \\
\hline 
\hline
\end{tabular}
\end{minipage}
\end{table}

\begin{figure}[H]
\centering
\vspace{5mm}
\includegraphics[width=0.7\textwidth]{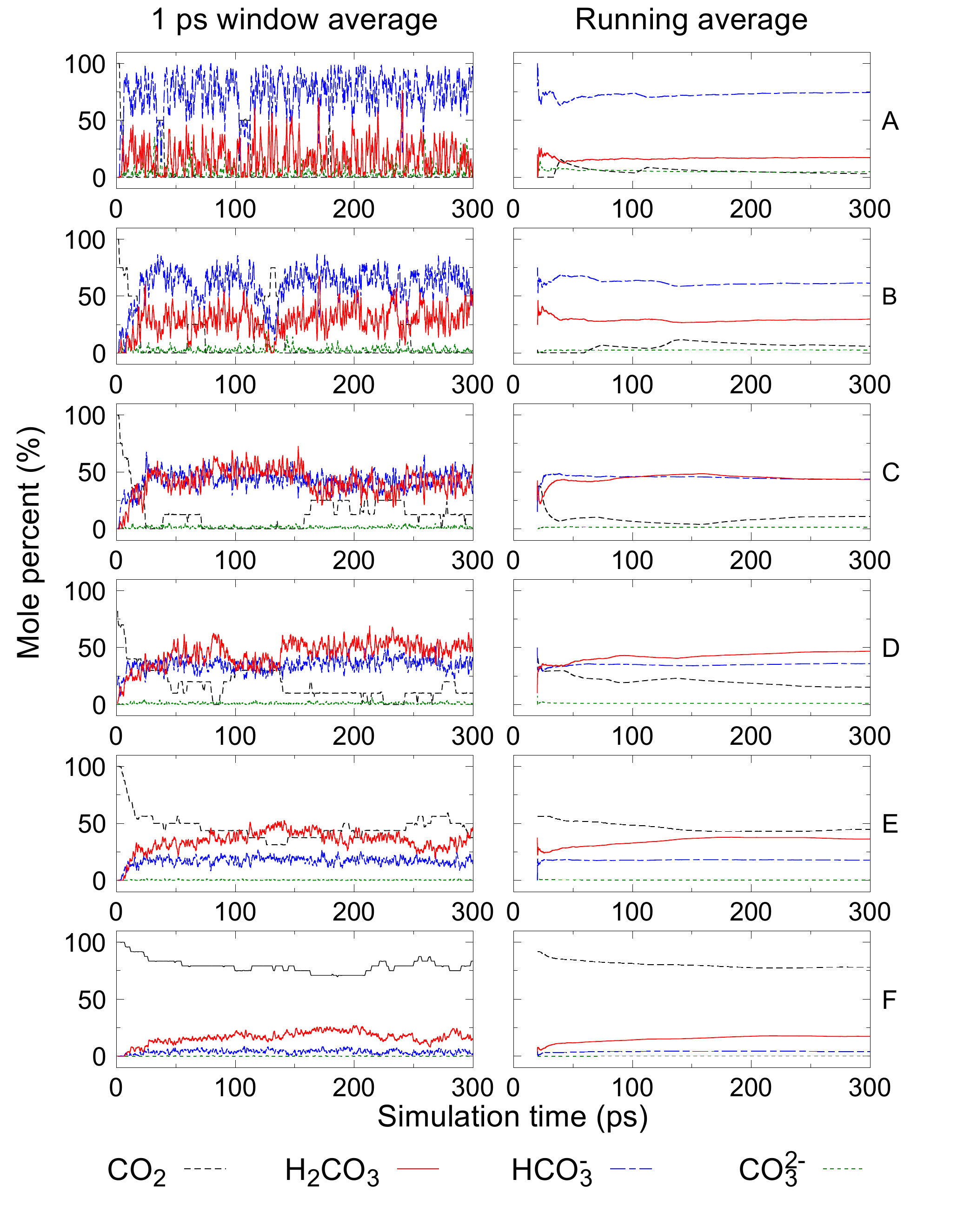}
\caption{Mole percents of carbon species as functions of simulation time in AIMD simulations. Left: 1-ps-window averages. Right: running averages. The initial mole fractions of CO$_2$(aq) (x(CO$_2$)) are (A) 0.032, (B) 0.067, (C) 0.143, (D) 0.185, (E) 0.333, and (F) 0.600. The temperature is 1000 K, and the pressure is $\sim$10 GPa.}
\label{running}
\end{figure}

\begin{figure}
\centering
\vspace{5mm}
\includegraphics[width=0.7\textwidth]{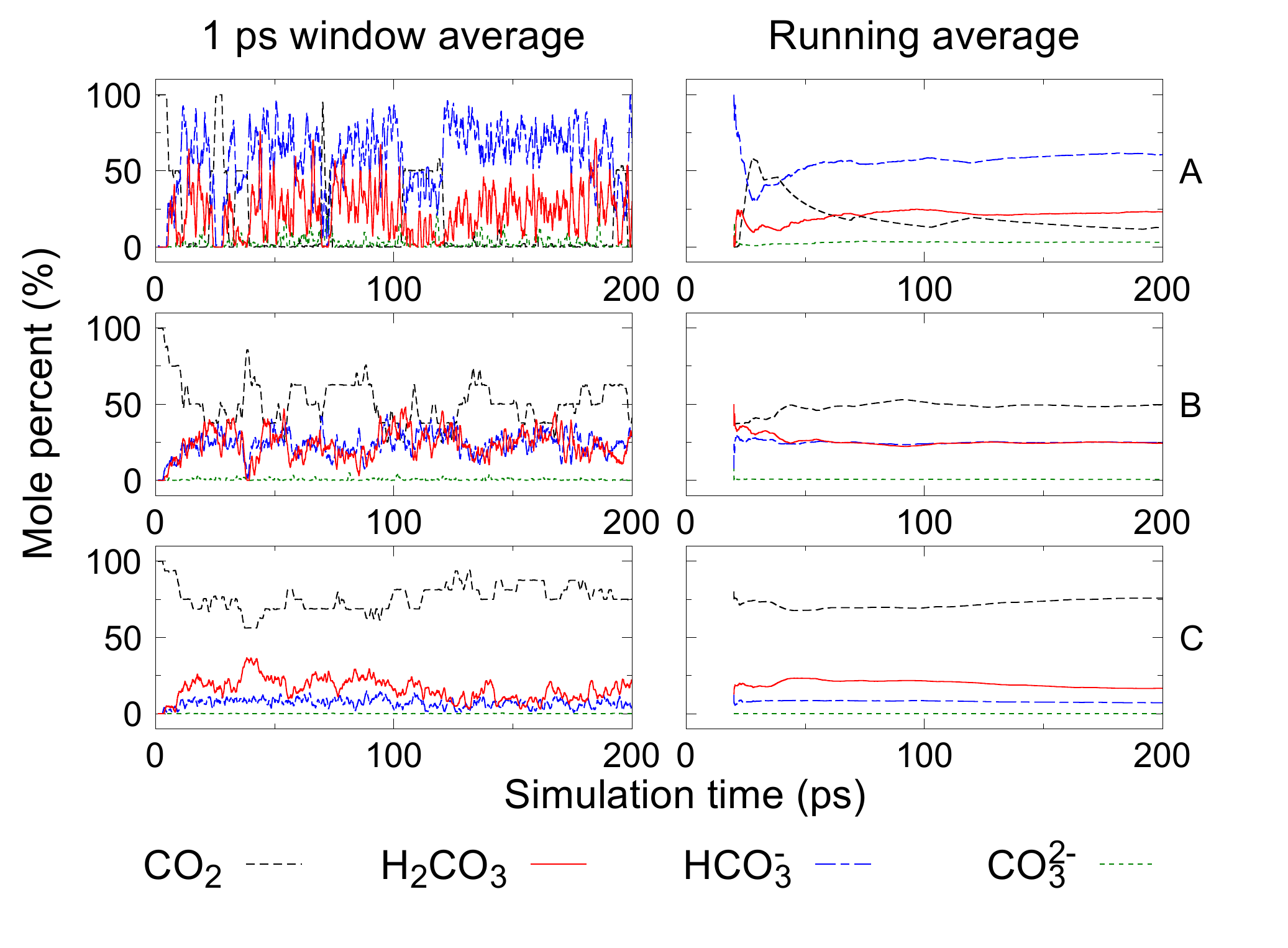}
\caption{
Mole percents of carbon species as functions of simulation time in AIMD simulations. Left: 1-ps-window averages. Right: running averages. The initial mole fractions of CO$_2$(aq) (x(CO$_2$)) are (A) 0.032, (B) 0.143, and (C) 0.333. The temperature is 1400 K, and the pressure is $\sim$10 GPa.}
\label{running1400}
\end{figure}

\begin{figure}
\centering
\vspace{5mm}
\includegraphics[width=0.7\textwidth]{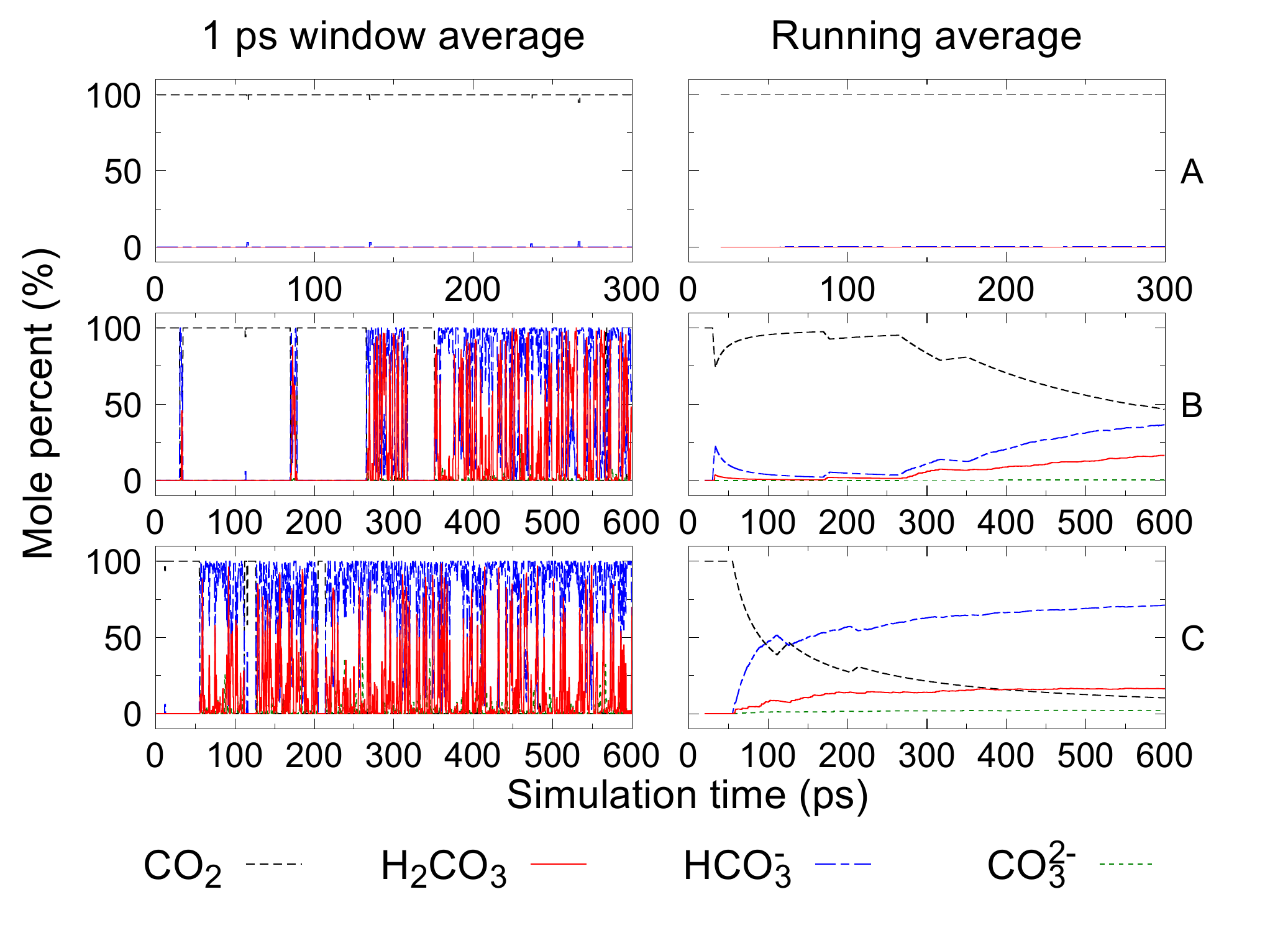}
\caption{
Mole percents of carbon species as functions of simulation time in AIMD simulations. Left: 1-ps-window averages. Right: running averages.  The pressures are (A) 3.8 GPa, (B) 6.1 GPa, and (C) 8.0 GPa. The temperature is 1000 K, and x(CO$_2$) is 0.016. 
}
\label{runningpressure}
\end{figure}

\begin{figure}
\centering
\vspace{5mm}
\includegraphics[width=0.7\textwidth]{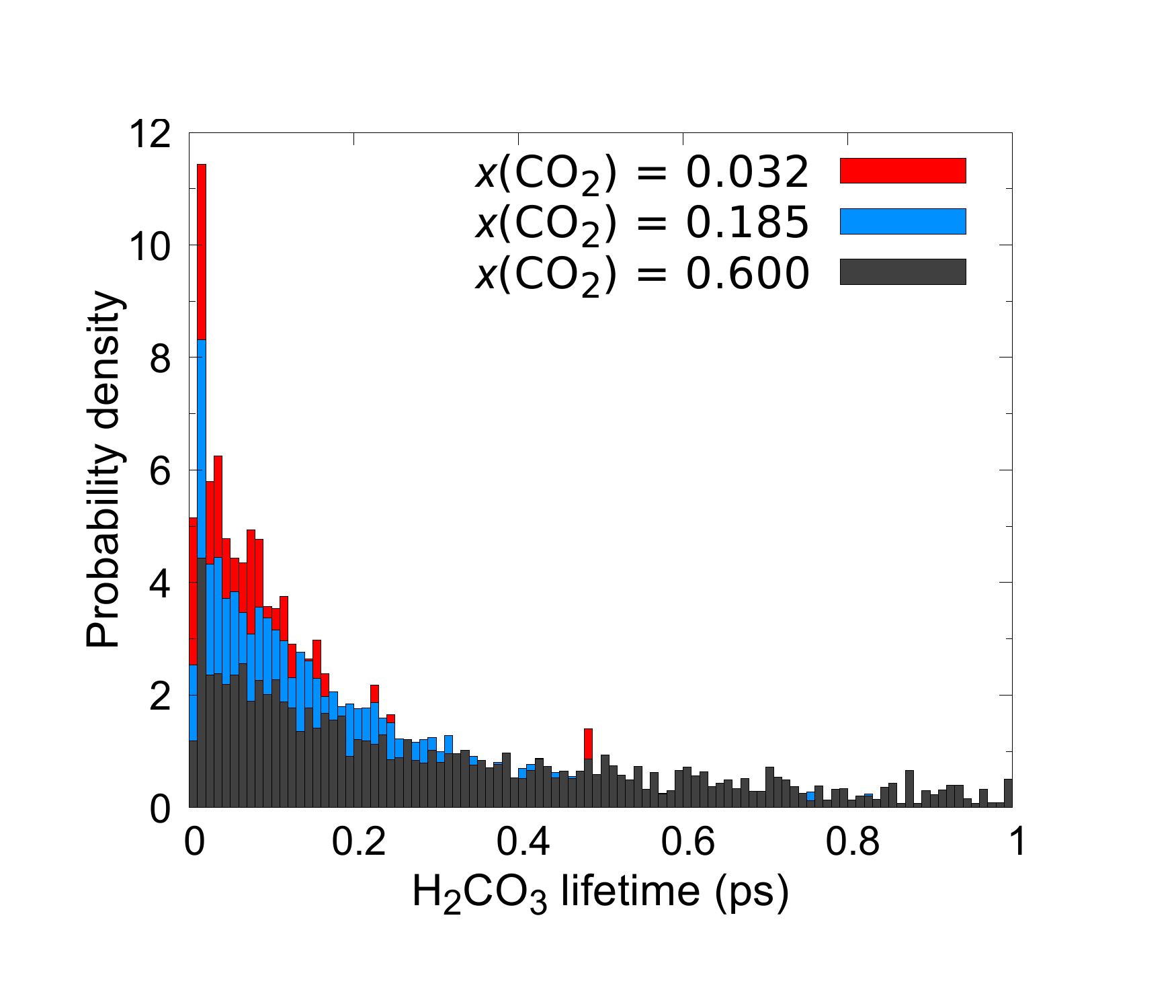}
\caption{Probability densities of lifetimes of H$_2$CO$_3$ at $\sim$10 GPa and 1000 K.
The initial mole fractions of CO$_2$(aq) (x(CO$_2$)) are 0.032, 0.185, and 0.600.}
\label{lifetimes}
\end{figure}

\begin{figure}
    \centering
    \includegraphics[width=1.0\textwidth]{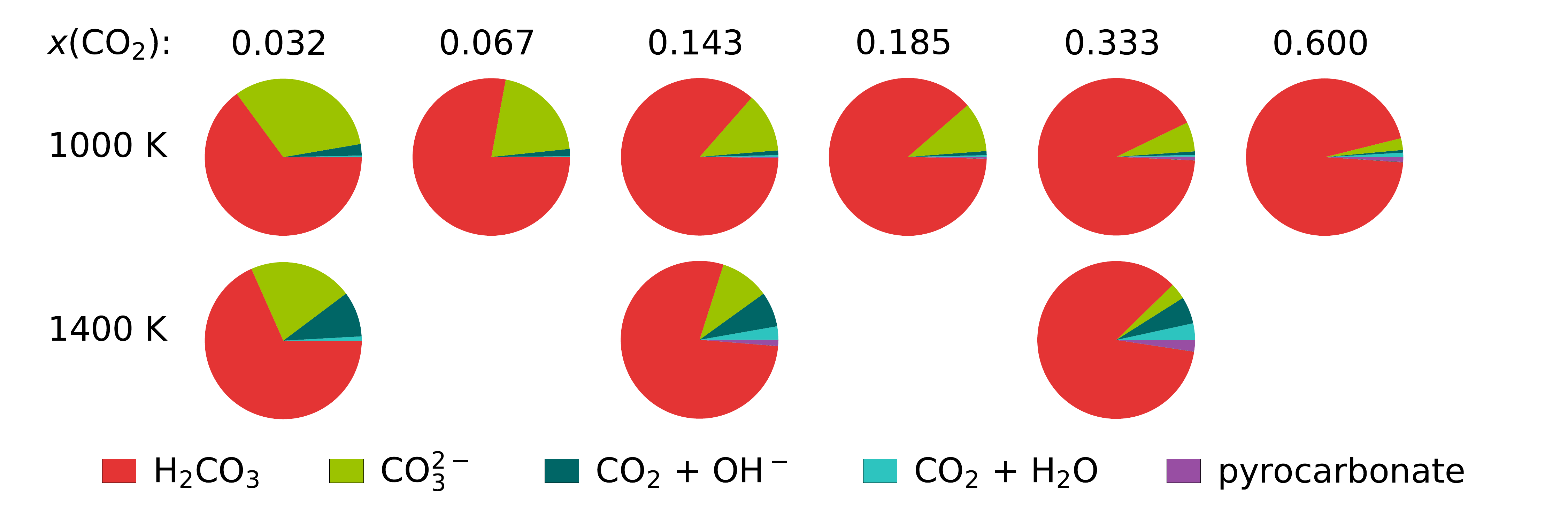}
    \caption{Compositions of reaction products of HCO$_3^-$ at $\sim$10 GPa. The reactions may be the decomposition (HCO$_3^-$ $\rightarrow$ CO$_2$ + OH$^-$), the protonation (HCO$_3^-$ + H$^+$ $\rightarrow$ H$_2$O + CO$_2$, HCO$_3^-$ + H$^+$ $\rightarrow$ H$_2$CO$_3$), the deprotonation, or the formation of C-O-C bonds.}
    \label{HCO3_reactions}
\end{figure}

\clearpage
\bibliography{references_SI}